\documentclass[a4paper,12pt]{article}

\usepackage{amsmath,amssymb,mathtools} 
\usepackage{color}
\usepackage{graphicx}
\usepackage{subfigure}
\usepackage{cite}           
\usepackage{hyperref}            
\usepackage{multirow,makecell}   
\usepackage{textcomp}
\usepackage{wasysym}
\usepackage{setspace}
\usepackage{verbatim}
\usepackage{float}
\usepackage{ulem}
\usepackage[utf8]{inputenc}
\usepackage[text={17.2cm,25.2cm},centering]{geometry}

\numberwithin{equation}{section}

\begin{document}

\title{\textbf{Holographic study of higher-order baryon number susceptibilities at finite temperature and density}}
	\author{Zhibin Li $^{1}$\footnote{lizhibin@zzu.edu.cn},
                Jingmin Liang $^{1}$\footnote{liangjingmin@gs.zzu.edu.cn},
                Song He $^{2,3}$\footnote{hesong@jlu.edu.cn (Corresponding author)}, and
                Li Li $^{4,5,6}$\footnote{ liliphy@itp.ac.cn (Corresponding author)}}
	\date{}
	
	\maketitle
	
	\vspace{-10mm}
	
	\begin{center}
		{\it
              $^{1}$ School of Physics and Microelectronics, Zhengzhou University, Zhengzhou 450001, China\\ \vspace{1mm}
		$^{2}$ Center for Theoretical Physics and College of Physics, Jilin University,
	Changchun 130012, China\\ \vspace{1mm}			
		$^{3}$ Max Planck Institute for Gravitational Physics (Albert Einstein Institute), Am Muhlenberg 1, 14476 Golm, Germany\\ \vspace{1mm}			
		$^{4}$ CAS Key Laboratory of Theoretical Physics, Institute of Theoretical Physics,
	Chinese Academy of Sciences, Beijing 100190, China\\ \vspace{1mm}			
		$^{5}$ School of Fundamental Physics and Mathematical Sciences,
	Hangzhou Institute for Advanced Study, University of Chinese Academy of Sciences, Hangzhou 310024, China\\ \vspace{1mm}			
		$^{6}$ Peng Huanwu Collaborative Center for Research and Education, Beihang University, Beijing 100191, China\\ \vspace{1mm} 
		}
		\vspace{10mm}
	\end{center}

	\begin{abstract}
The cumulants of baryon number fluctuations serve as a good probe for experimentally exploring the QCD phase diagram at finite density, giving rise to characteristic fluctuation patterns associated with a possible critical endpoint (CEP). 
 We compute the higher-order baryon number susceptibilities at finite temperature and baryon chemical potential using a holographic QCD model to address the non-perturbative aspect of strongly coupled QCD matter. The model
 can accurately confront lattice QCD data on a quantitative level and the location of the CEP is found to fall within the range accessible to upcoming experimental measurements. The baryon number susceptibilities up to the twelfth order are computed, and the collision energy dependence of different ratios of these susceptibilities is examined along the chemical freeze-out line. The holographic results show quantitative agreement with experimental data and the functional renormalization group results in a large collision energy range, with all ratios exhibiting a peak structure around 5-10 GeV. The mismatching between our holographic results with experimental data for sufficiently low collision energy is possibly due to non-equilibrium effects and complex experimental environments. The future experiments with measurements in the low collision energy range $\sqrt{S_{NN}}\approx 1-10~\text{GeV}$ and reduced experimental uncertainty could reveal more non-monotonic behavior signals which can be used to locate the CEP.

	\end{abstract}
	
	\baselineskip 18pt
	\thispagestyle{empty}
	\newpage
	
	\tableofcontents

\maketitle
%%%%%%%%%%%%%%%%%%%%%%%%%%%%%%%%%%%%%%%%%%%%%%
\section{Introduction}\label{sec:01}
%%%%%%%%%%%%%%%%%%%%%%%%%%%%%%%%%%%%%%%%%%%%%%

Obtaining a quantitative understanding of the QCD phase diagram at finite temperature $T$ and baryon chemical potential $\mu_B$ remains remarkably challenging due to the strongly coupled nature of the system under extreme conditions. Significant efforts have been devoted to this problem over the past few decades. Lattice QCD, formulated on a grid of points in space and time, provides reliable information from first principles at small $\mu_B$ where the sign problem does not hinder numerical calculations. Lattice QCD calculations indicate that the chiral and confinement/deconfinement phase transitions likely occur as an analytic crossover for small $\mu_B$, with mixing of the transitions~\cite {Borsanyi:2010cj, Borsanyi:2013bia, HotQCD:2014kol}. On the other hand, several effective theories, including the Dyson-Schwinger equation (DSE)~\cite{Xin:2014ela, Gao:2016qkh,Qin:2010nq,Shi:2014zpa,Fischer:2014ata,Gao:2020qsj}, the Nambu-Jona-Lasinio (NJL) model~\cite{Asakawa:1989bq,Schwarz:1999dj,Li:2018ygx,Zhuang:2000ub}, and the functional renormalization group (FRG)~\cite{Fu:2019hdw,Zhang:2017icm,Fu:2021oaw}, suggest the existence of a first-order phase transition at large $\mu_B$, which would terminate at a critical point known as the QCD critical endpoint (CEP). However, the exact location of the CEP is still a matter of debate, with no conclusive constraints from any model calculations thus far. Nevertheless, lattice QCD results disfavor the existence of the CEP for $\mu_B/T\le 3$ and $\mu_B<300~\text{MeV}$~\cite{Vovchenko:2017gkg,Borsanyi:2020fev,Bazavov:2020bjn,Borsanyi:2021sxv,Bollweg:2022fqq,Philipsen:2021qji}.

The critical physics associated with the CEP is expected to have a localized impact in its vicinity~\cite{Schaefer:2006ds}. Numerous theoretical studies have revealed intriguing non-trivial patterns in the ratios of conserved charge distributions $C_n$ around the CEP~\cite{Asakawa:2009aj, Schaefer:2011ex, Fan:2016ovc, Portillo:2016fso, Fan:2017mrk, Li:2017ple, Li:2018ygx, Fu:2021oaw,Zhao:2023xpj}.  Non-monotonic variations of conserved charge fluctuations with respect to the $T$ and $\mu_B$ along the phase boundary could arise from critical physics in the vicinity of a CEP which can serve as signals for CEP~\cite{Stephanov:2011pb}. Experimental measurement has suggested a significant overlap between the chemical freeze-out region and the crossover region for $\mu_B/T\le 3$~\cite{STAR:2017sal}, implying that the freeze-out line is likely to pass through the vicinity of the CEP, provided that the CEP is not far beyond $\mu_B/T\approx 3$. This indicates that $C_n$ and their ratios along the chemical freeze-out line may take similar non-monotonic behavior~\cite{Fan:2016ovc, Fan:2017mrk, Li:2017ple, Portillo:2017gfk, Li:2018ygx, Wang:2018sur, Fu:2021oaw, Zhao:2023xpj, Huang:2023ogw}. Remarkably, these ratios can be directly linked to measurable quantities in experiments, such as the mean, variance, skewness, and kurtosis, making it feasible to locate the CEP by measuring cumulants of conserved charge distributions in experimental studies.

Indeed, the potential existence of the CEP and the first-order phase transition has motivated dedicated experimental programs, particularly in relativistic heavy-ion collisions. In heavy ion collisions, the early non-equilibrium state of quarks and gluons will become the final hadronic states after a chemical freeze-out. Moreover, regions of large $\mu_B$ can be experimentally researched by lowering the beam energy. 
In recent years, relativistic heavy-ion collision experiments have made significant progress in the search for the CEP~\cite{Luo:2017faz, STAR:2021iop}. Various cumulants, including net-proton, net-charge, and net-kaon cumulants, have been measured at different collision energies. Notably, recent STAR data on net-proton distributions $\kappa\sigma^2$ in Au+Au collisions as a function of collision energy $\sqrt{S_{NN}}$ shows a non-monotonic variation, exhibiting a peak structure near $\sqrt{S_{NN}}\approx 7~\text{GeV}$, which could be an experimental signature of the CEP~\cite{STAR:2020tga, STAR:2021fge, Pandav:2022xxx, Zhang:2022evi}. Furthermore, measurements have been extended to higher-order cumulants, including the sixth-order~\cite{STAR:2022vlo} and eighth-order~\cite{Pandav:2023lis} cumulants of net-proton fluctuations. Non-monotonic dependencies on collision energy have also been observed in the fifth-order and sixth-order cumulant data of net-proton fluctuations in $0-40\%$ centrality Au-Au collisions~\cite{STAR:2022vlo}.

Given the limited experimental data available that is confined to the crossover region with $\mu_B/T\leq 3$, it is crucial to deepen our understanding of conserved charge fluctuations at high $\mu_B$, where lattice simulations face challenges due to the sign problem. To address this non-perturbative aspect, we employ holographic duality to map the strongly correlated physics of the QCD phase diagram to a higher-dimensional gravity system. Holography offers a convenient framework to incorporate real-time dynamics and study transport properties at finite temperatures and densities. Our holographic model has been demonstrated to capture the essential characteristics of realistic QCD and successfully confront lattice QCD data with 2+1 flavors on a quantitative level~\cite{Cai:2022omk}. We have constructed the phase diagram in terms of $T$ and $\mu_B$, and determined the location of the CEP at $(T_C=105~\text{MeV}, \mu_C=555~\text{MeV})$ which falls within the range accessible to upcoming experimental measurements~\cite{Cai:2022omk}. In this study, we shall investigate the behavior of baryon number fluctuations over a wide range of temperatures and baryon chemical potentials. We will compare
our holographic results with the experimental measurements and will provide further theoretical predictions.~\footnote{Previous studies on the baryon susceptibilities using holographic QCD can be found in~\cite{Rougemont:2015ona, Portillo:2017gfk, Critelli:2017oub}. The model was fixed by matching at zero baryon chemical potential to the lattice equation of state from~\cite{Borsanyi:2013bia}. It predicts a CEP at a significantly different location in the phase diagram from our model~\cite{Cai:2022omk}.}

The rest of this paper is organized as follows. In section~\ref{sec:02}, we briefly review our holographic QCD model, including optimized parameters and equations of states. Section~\ref{sec:03} shows the baryon number susceptibilities up to the twelfth order at $\mu_B=0$. We also compare our holographic results with the lattice QCD data. Section~\ref{sec:04} compares our results to available experimental data from heavy ion collisions along the chemical freeze-out line. We will present the prediction for the beam energy dependence of baryon number susceptibilities for which no experimental data is available yet. We conclude with some discussion in Section~\ref{sec:05}.

%%%%%%%%%%%%%%%%%%%%%%%%%%%%%%%%%%%%%%%%%%%%%%
\section{Holographic QCD model}\label{sec:02}
%%%%%%%%%%%%%%%%%%%%%%%%%%%%%%%%%%%%%%%%%%%%%%

We consider the $2+1$ flavor holographic QCD model established in~\cite{Cai:2022omk}. The gravitational action takes the following form.
\begin{equation}\label{eq21}
  S_M =\frac{1}{2\kappa_N^2}\int d^5x\sqrt{-g}[\mathcal{R}-\frac{1}{2}\nabla_\mu\phi\nabla^\mu\phi-\frac{Z(\phi)}{4}F_{\mu\nu}F^{\mu\nu}-V(\phi)]\,,
\end{equation}
with $g_{\mu\nu}$ the metric of the bulk spacetime, $\phi$ the scalar field, and $A_\mu$ the gauge field incorporating finite baryon chemical potential and baryon density. Here $V(\phi)$ and $Z(\phi)$ are two free couplings in our bottom-up model. The non-perturbative effects and flavor dynamics are effectively adopted into the model parameters by matching up-to-date lattice QCD data.~\footnote{This approach involving a bulk nonconformal dilatonic scalar and a $U(1)$ gauge field has been widely used in holographic QCD, see \emph{e.g.}~\cite{DeWolfe:2010he, DeWolfe:2011ts, Cai:2012xh, Cai:2012eh, Finazzo:2013efa, Yang:2014bqa, Critelli:2017oub, Li:2017ple, Chen:2017cyc, Knaute:2017opk, Fang:2018axm, Ballon-Bayona:2020xls, Li:2020hau, Grefa:2021qvt, He:2022amv, Grefa:2022fpu}.}  

The bulk spacetime metric with matter fields $\phi$ and $A_\mu$ reads
\begin{align}\label{eq23}
  ds^2 &=-e^{-\eta(r)}f(r)dt^2+\frac{dr^2}{f(r)}+r^2(d x_1^2+d x_2^2+d x_3^2)\,,\nonumber\\
  \phi &=\phi(r),\quad\quad\quad A_\mu dx^\mu=A_t(r)dt\,,
\end{align}
where $r$ is the holographic radial coordinate for which $r\rightarrow \infty$ corresponds to the AdS boundary. Denoting the location of the event horizon as $r=r_h$ where $f(r_h)=0$, the Hawking temperature and the entropy density are given by
\begin{equation}\label{eq24}
  T=\frac{1}{4\pi}f'(r_h)e^{-\eta(r_h)/2},  \quad  s=\frac{2\pi}{\kappa_N^2}r_h^3.
\end{equation}
Substituting~\eqref{eq23} into the action~\eqref{eq21}, one obtains the equations of motion that have to be solved numerically to obtain the hairy black holes. Then, the related thermodynamic quantities, including the energy density $\mathcal{E}$, the pressure $P$, and the baryon chemical potential $\mu_B$ can be obtained using holographic renormalization (see~\cite{Cai:2022omk} for more technical details).

\begin{figure}[htbp]
\centering
\includegraphics[width=.45\textwidth]{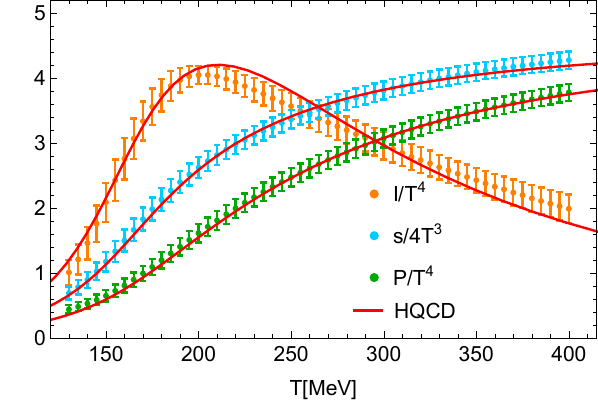}
\qquad
\includegraphics[width=.45\textwidth]{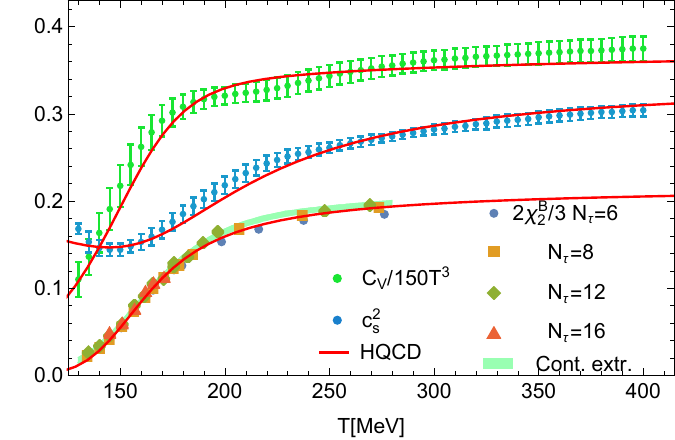}
\caption{Thermodynamics at $\mu_B=0$ from lattice QCD results~\cite{HotQCD:2012fhj, HotQCD:2014kol, Bazavov:2017dus} compared to our holographic model (red solid curves). \textbf{Left panel}: the entropy density $s$, the pressure $P$, and the trace anomaly $I=\mathcal{E}-3P$. \textbf{Right panel}: the specific heat $C_V$, the squared speed of sound $c_s^2$, and the baryon susceptibility $\chi_B^2$.}\label{fig1}
\end{figure}

The two couplings in~\eqref{eq21} are parameterized to be~\cite{Cai:2022omk}
\begin{equation}\label{eq22}
\begin{split}
 V(\phi)&=-12 \cosh\left[c_1\phi\right]+\left(6c_1^2-\frac{3}{2}\right)\phi^2+c_2\phi^6\,, \\
Z(\phi)&=\frac{1}{1+c_3}{\mathrm sech}[c_4\phi^3]+\frac{c_3}{1+c_3}e^{-c_5\phi}\,,
\end{split}
\end{equation}
where $c_1$ to $c_5$ are free parameters.
The other two free parameters are the effective Newton constant $\kappa_N^2$ and a characteristic energy scale set by the leading source term of $\phi$, \emph{i.e.} $\phi_s=\lim_{r\rightarrow\infty} r\phi$. The latter breaks the scale invariance of the boundary system to essentially describe the QCD dynamics as there is no conformal symmetry in real QCD.
All the above parameters are fixed completely by fitting the lattice QCD data at zero net-baryon density~\cite{HotQCD:2012fhj, HotQCD:2014kol, Bazavov:2017dus} and their values are summarized in Table~\ref{table1}. The parameter $b$ is from the holographic renormalization and is necessary to satisfy the lattice QCD simulation at $\mu_B=0$. We have made a slight modification to the value of $c_4$ to enhance the agreement with lattice data compared to the previous setup~\cite{Cai:2022omk}. Nevertheless, this modification yields almost the same  location of the CEP as the one of~\cite{Cai:2022omk}. We compare various thermodynamic quantities from our holographic setup with lattice simulation in Fig.~\ref{fig1}. One case sees that the temperature dependence of all those quantities agrees well with lattice QCD with $2+1$ flavors~\cite{HotQCD:2012fhj, HotQCD:2014kol, Bazavov:2017dus}.
\begin{table}[htbp]
\centering
\begin{tabular}{|c|c|c|c|c|c|c|c|c|}
\hline
    model  &  $c_1$  &  $c_2$  & $c_3$ & $c_4$  & $c_5$ & $\kappa_N^2$ & $\phi_s $ [MeV] & b \\ \hline
    2+1 flavor  &  0.710  &  0.0037  & 1.935  & 0.091  & 30 & $2\pi(1.68)$  &  1085 & -0.27341 \\ \hline
\end{tabular}
\caption{Parameters for our $2+1$ flavor QCD model by matching the lattice simulation.}
    \label{table1}
\end{table}
%%%

The generalized susceptibilities are closely related to various cumulants of the baryon number distribution measured in heavy-ion collision experiments. Here we focus on the susceptibilities of the baryon number $\chi_n^B$ that are defined through the $n$-th order derivatives of the pressure w.r.t. the baryon chemical potential.
\begin{equation}\label{eq26}
  \chi_n^B(T,\mu_B)=\frac{\partial^n}{\partial (\mu_B/T)^n}\frac{P}{T^4}\,.
\end{equation}
For example, to apply the QCD simulation to a finite density case, one could consider a Taylor expansion in baryon chemical potential
that underlies the extension of lattice results that are only available at $\mu_B=0$. 
\begin{align}\label{eq25}
\frac{P\left(T,\mu_B\right)-P\left(T,0\right)}{T^4}&=\sum_{n=1}^{\inf}\frac{\chi_{2n}^B\left(T\right)}{\left(2n!\right)}\left(\frac{\mu_B}{T}\right)^{2n} \nonumber\\
 =\frac{1}{2}\chi_2^B\left(T\right)\hat{\mu}_B^2&\left(1+\frac{1}{12}\frac{\chi_4^B\left(T\right)}{\chi_2^B\left(T\right)}\hat{\mu}_B^2+\frac{1}{360}\frac{\chi_6^B\left(T\right)}{\chi_2^B\left(T\right)}\hat{\mu}_B^4+\frac{1}{20160}\frac{\chi_8^B\left(T\right)}{\chi_2^B\left(T\right)}\hat{\mu}_B^6+\dots\right),
\end{align}
where $\hat{\mu}_B=\mu_B/T$ is the reduced baryon chemical potential and $\chi_{2n}^B(T)$ are baryon number susceptibilities at $\mu_B=0$. Note that $\chi_{2n+1}^B(T,\mu_B=0)=0$ due to the CP symmetry.

The corresponding cumulant of baryon distribution is given by
\begin{equation}\label{eq27}
  C_n^B=V T^3 \chi_n^B\,,
\end{equation}
with $V$ the freeze-out volume in heavy-ion collisions. The ratios of these cumulants cancel out volume dependence and are observable quantities in experiments. In particular, the skewness $S_B$ and the kurtosis $\kappa_B$ of baryon distribution are given by
\begin{equation}\label{eq28}
  S_B=\frac{C_3^B}{\left(\sigma_2^B\right)^{3/2}}\hspace{0.4cm} \text{and} \hspace{0.4cm}  \kappa_B=\frac{C_4^B}{\left(\sigma_2^B\right)^{2}}\,,
\end{equation}
with the notation $M_B=C_1^B$ for the mean and $\sigma_2^B=C_2^B$ for the variance.

\begin{figure}[htbp]
\centering
\includegraphics[width=.45\textwidth]{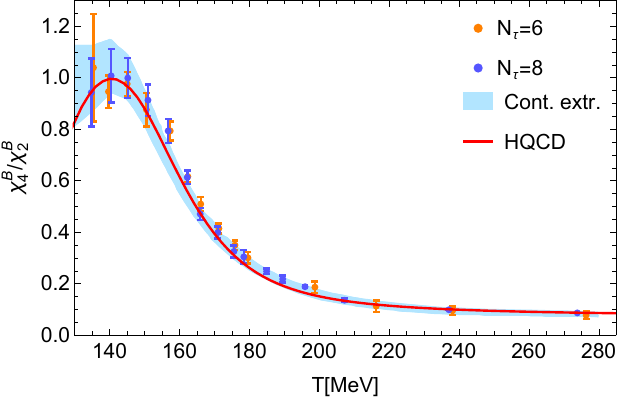}
\qquad
\includegraphics[width=.45\textwidth]{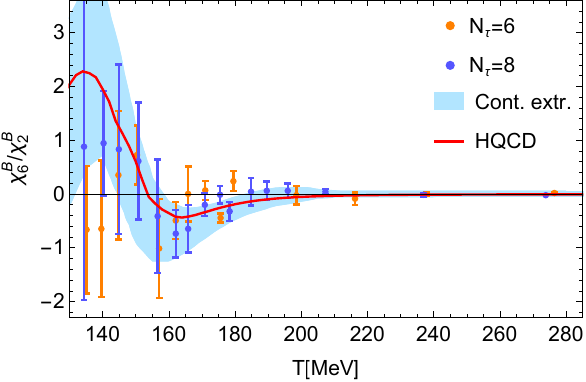}\\
\includegraphics[width=.47\textwidth]{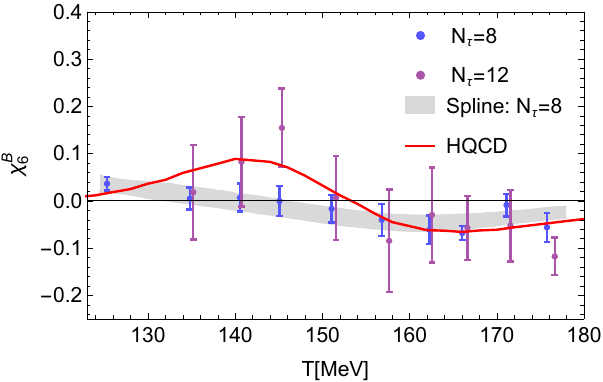}
\qquad
\includegraphics[width=.45\textwidth]{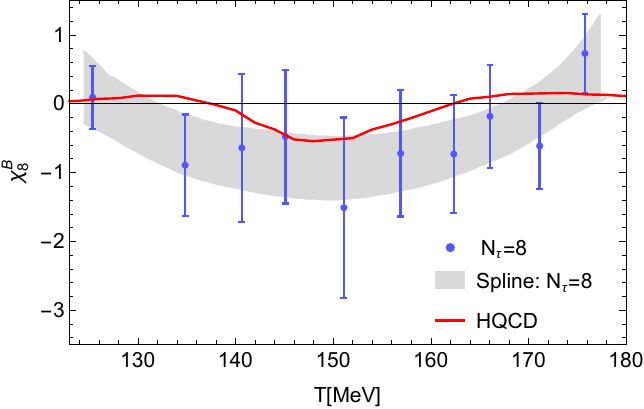}\\
  \caption{Baryon number susceptibilities $\chi^B_4/\chi^B_2$ (top left), $\chi^B_6/\chi^B_2$ (top right), $\chi^B_6$ (bottom left) and $\chi^B_8$ (bottom right) at $\mu_B=0$ compare with lattice data~\cite{Bazavov:2017dus,Bazavov:2020bjn,Bollweg:2022rps}. The light-blue band denotes the region of continuous extrapolation from lattice QCD simulation.}\label{fig2}
\end{figure}
%
%%%%%%%%%%%%%%%%%%%%%%%%%%%%%%%%%%%%%%%%%%%%%%
\section{Baryon number susceptibility at $\mu_B=0$}\label{sec:03}
%%%%%%%%%%%%%%%%%%%%%%%%%%%%%%%%%%%%%%%%%%%%%%
As a benchmark test, we present the numerical results of the baryon number susceptibilities at $\mu_B=0$ and compare them with available lattice QCD simulation. 

The behavior of $\chi_2^B$ has been depicted in the right panel of Fig.~\ref{fig1}. One can see clearly that the holographic result is in good agreement with the latest lattice data from HotQCD group~\cite{Bazavov:2017dus}.
We compare higher-order susceptibilities in Fig.~\ref{fig2}. The top two plots are for $\chi_4^B/\chi_2^B$ and $\chi_6^B/\chi_2^B$, respectively. The points with error bars are the lattice data with different $N_\tau$, and the light-blue band represents the region of continuous extrapolation based on the lattice data~\cite{Bazavov:2017dus,Bazavov:2020bjn,Bollweg:2022rps}. The holographic QCD results are given by solid red curves. Our direct computation matches well with the lattice data, particularly the results of the continuous extrapolation. 
The temperature dependence of $\chi_6^B$ and $\chi_8^B$ is presented in the bottom two plots of Fig.~\ref{fig2}. Given that the uncertainty of the HotQCD data is relatively high. It is still challenging to analyze the behavior of $\chi_8^B$. Nevertheless, our $\chi_8^B$ results qualitatively agree with those from the W-B data~\cite{Borsanyi:2018grb} and the FRG result~\cite{Fu:2021oaw}.

While $\chi_2^B$ monotonically increases with the temperature at $\mu_B=0$, higher-order susceptibilities versus temperature yield more complicated behaviors (see Fig.~\ref{fig2}). The ratio $\chi_4^B/\chi_2^B$ remains positive throughout but it initially increases to a peak at $T\approx 140~\text{MeV}$, before decreasing to approximately $0.1$ at high temperatures. Similarly, $\chi_6^B/\chi_2^B$ exhibits an increase to about $2.2$ at $T\approx 140~\text{MeV}$, followed by a decrease to $-0.4$ at $T\approx 165~\text{MeV}$ and a subsequent increase to around $0$. Thus, $\chi_6^B/\chi_2^B$ displays a peak and a dip. The behavior of $\chi_6^B$ is similar to that of $\chi_6^B/\chi_2^B$ due to the monotonically increasing dependence of $\chi_2^B$ on temperature. For $\chi_8^B$, our holographic result suggests that it initially grows from $0$ to approximately $0.2$ by increasing $T$, then decreases to around $-0.5$, and subsequently increases to $0.2$ again at high temperatures. 

\begin{figure}[htbp]
\centering
\includegraphics[width=.465\textwidth]{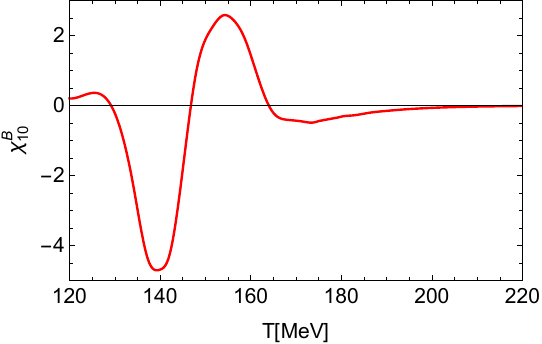}
\qquad
\includegraphics[width=.475\textwidth]{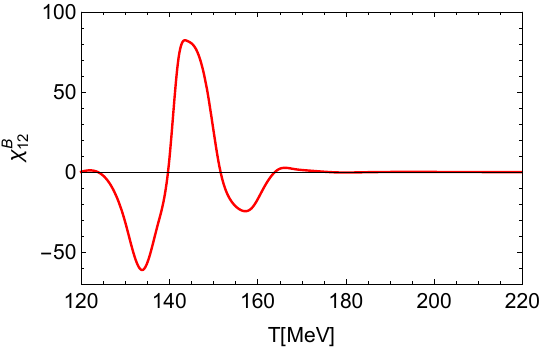}
  \caption{The temperature dependence of higher order baryon number susceptibilities $\chi^B_{10}$ (left) and $\chi^B_{12}$ (right) from our holographic theory at $\mu_B=0$.}\label{fig3}
\end{figure}
There is no available lattice data for higher-order $\chi_n^B$ with $n\geq 10$ thus far. In Fig.~\ref{fig3}, we show our theoretical computation for $\chi^B_{10}$ (left) and $\chi^B_{12}$ (right) as a function of temperature. Both cases display a more intricate behavior with temperature, featuring both increasing and decreasing trends, as well as positive and negative values. It will be of great interest to compare our results with future lattice QCD data for $\chi^B_{10}$ and $\chi^B_{12}$, which would allow for a quantitative assessment of the accuracy of our model.

%%%%%%%%%%%%%%%%%%%%%%%%%%%%%%%%%%%%%%%%%%%%%%
\section{Baryon number susceptibility along chemical freeze-out line}\label{sec:04}
%%%%%%%%%%%%%%%%%%%%%%%%%%%%%%%%%%%%%%%%%%%%%%

In this section, we begin by examining the ratios of various higher-order baryon number susceptibilities at the collision energies measured by RHIC, following the chemical freeze-out data determined by Gupta et al.~\cite{Gupta:2020pjd}. We will consider two fitted chemical freeze-out lines, aiming to capture the key characteristics of the CEP in the $T-\mu_B$ plane. As we will show, our theoretical predictions demonstrate some quantitative agreement with the most recent experimental data~\cite{Xu:2016mqs, STAR:2020tga, STAR:2021fge, STAR:2022vlo, Pandav:2023lis} as well as the results obtained from the FRG approach~\cite{Fu:2021oaw}. 
%The outcomes of our analysis are presented in Fig.~\ref{fig4}, Fig.~\ref{fig6}, and Fig.~\ref{fig7}.

%
\begin{figure}[htbp]
\centering
\includegraphics[width=.465\textwidth]{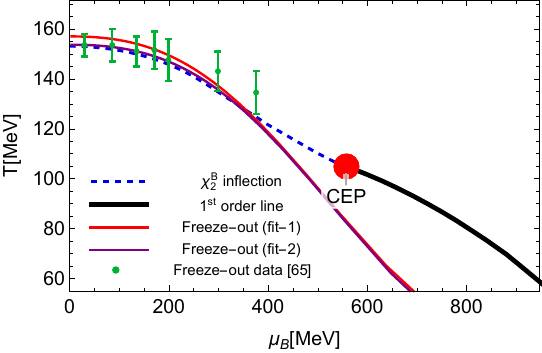}
\qquad
\includegraphics[width=.45\textwidth]{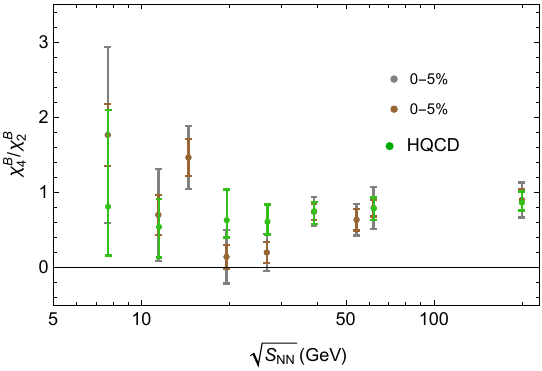}
  \caption{ {\textbf{Left panel}: The phase diagram of our holographic QCD model, where the blue dashed line corresponds to the crossover line determined by $\chi^B_2$ inflection. The CEP is shown as a bold red dot. The thick black line represents the first-order transition line, while the red and purple lines depict the two fitted chemical freezeout lines utilized in this study. The green data, accompanied by errors, corresponds to the chemical freezeout data extracted from~\cite{Gupta:2020pjd}.}  \textbf{Right panel}: $\chi_4^B/\chi_2^B$ along the chemical freeze-out line compares with STAR data of net-proton distributions in $0-5\%$ centrality Au-Au collisions~\cite{STAR:2021fge}. The grey and brown error bars represent the statistical and systematic uncertainties, respectively. We have taken the position of chemical freeze-out directly from the HRG model~\cite{Gupta:2020pjd}. }\label{fig4}
\end{figure}

{The phase diagram of our holographic QCD model is presented in the left panel of Fig.~\ref{fig4}, wherein the blue dashed line corresponds to the crossover line determined by $\chi^B_2$ inflection. The location of CEP is marked by a bold red dot. The thick black line represents the first-order transition line, while the red and purple lines depict the two chemical freeze-out lines fitted in our present work (see more details around Fig.~\ref{fig5} below). The green data, accompanied by errors, corresponds to the chemical freeze-out data given in the hadron resonance gas (HRG) model~\cite{Gupta:2020pjd}.} In the right panel of Fig.~\ref{fig4}, we present a comparison of the ratio of the fourth-order to second-order baryon number susceptibilities, $\chi_4^B/\chi_2^B$, obtained using our holographic QCD model for collision energies of $\sqrt{S_{NN}}=7.7$, $11.5$, $19.6$, $27$, $39$, $62.4$, and $200~\text{GeV}$, with experimental data of net-proton distributions for $0-5\%$ centrality Au-Au collisions from STAR~\cite{STAR:2021fge}. The grey and brown error bars represent the statistical and systematic uncertainties of the experimental data points, respectively. The fitted positions on the $T-\mu_B$ phase diagram for various collision energies have been obtained from~\cite{Gupta:2020pjd}. It is worth noting that the uncertainty in our theoretical results is attributed to the imprecision in determining the location of the fixed collision energy on the $T-\mu_B$ phase diagram~\cite{Gupta:2020pjd}. The direct comparison of Fig.~\ref{fig4}, without any adjustable parameters, demonstrates a significant overlap between our theoretical results and experimental data at different collision energies.

Different centrality in heavy-ion collisions corresponds to different positions of chemical freeze-out in the $T-\mu_B$ phase diagram~\cite{STAR:2017sal}. Therefore, one should choose an appropriate freeze-out line for a given centrality~\cite{Mukherjee:2023ijv}. In the present study, we consider two chemical freeze-out lines by fitting the $dN/dy$ and $4\pi$ yields data in~\cite{Andronic:2009jd,Andronic:2017pug}. Also note that our fitting procedure does not yield a unique ``optimal" fit for the data in~\cite{Andronic:2009jd,Andronic:2017pug}, as we aim to optimize both the range of experimentally fitted freeze-out data and the degree of agreement between theoretical values of $\chi_m^B/\chi_n^B$ and experimental data.
\begin{figure}[htbp]
\centering
\includegraphics[width=.47\textwidth]{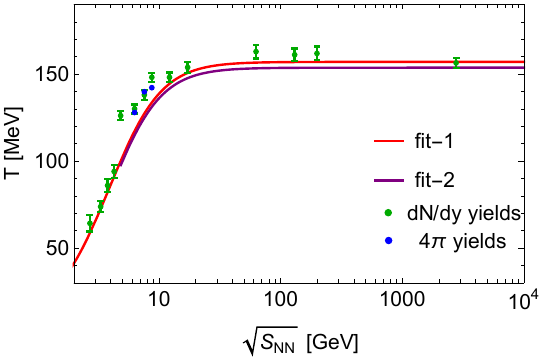}
\qquad
\includegraphics[width=.47\textwidth]{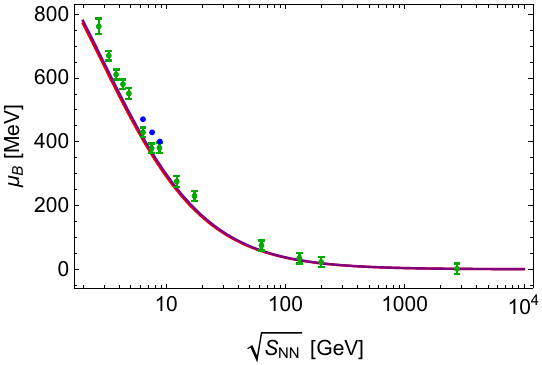}
  \caption{The two chemical freeze-out lines we used in this work by fitting the $dN/dy$ and $4\pi $ yields data~\cite{Andronic:2009jd, Andronic:2017pug}.The locations of the two fitted chemical freeze-out lines can be observed in the left panel of Figure~\ref{fig4}, within the context of the phase diagram. \textbf{Left panel}: The temperature as a function of the collision energy. \textbf{Right panel}: The baryon chemical potential versus the collision energy.}\label{fig5}
\end{figure}
Fig.~\ref{fig5} displays the chemical freeze-out lines corresponding to centrality ranges of $0-5\%$ (red line) and $0-40\%$ (purple line), respectively. The fitting formula for the two chemical freeze-out lines is given by
\begin{equation}\label{eq41}
  \mu_B=\frac{a}{1+b \sqrt{S_{NN}}},\hspace{1cm} T=\frac{T_{lim}}{1+exp\left[ c-\ln{\left(d \sqrt{S_{NN}}+e\right)/0.45 } \right]}\,,
\end{equation}
with $a, b, c, d, e, T_{lim}$ free parameters.
The corresponding parameters used to fit both lines of Fig.~\ref{fig5} are summarized in Table~\ref{table2}. {The resulting freeze-out lines are shown in the $T-\mu_B$ phase diagram for both cases in the right panel of Fig.~\ref{fig4}.}
\begin{table}[htbp]
\centering
\begin{tabular}{|c|c|c|c|c|c|c|}
\hline
   &  $a$ [MeV]  &  $b$ & $T_{lim}$ [MeV]  & $c$ & $d$ & $e$ \\ \hline
    fit-1  &  1307.5  &  0.35 & 157.0  & 3.25 & 1  & 0.7 \\ \hline
    fit-2  &  1307.5  &  0.34 & 153.6  & 3.41 & 1.1 & 0.6 \\ 
    \hline
\end{tabular}
\caption{Parameters for two chemical freeze-out lines in Fig.~\ref{fig5} by matching the $dN/dy$ and $4\pi $ yields data in~\cite{Andronic:2009jd,Andronic:2017pug}.}
    \label{table2}
\end{table}
\begin{figure}[htbp]
\centering
\includegraphics[width=.47\textwidth]{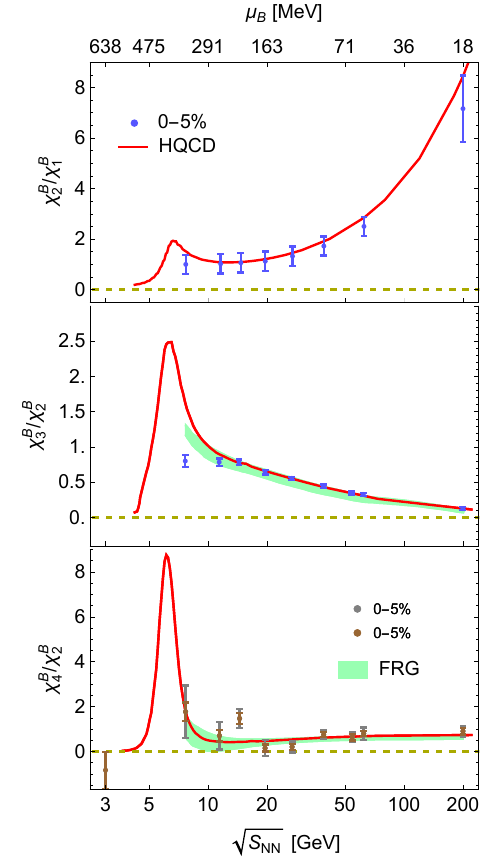}
\qquad
\includegraphics[width=.47\textwidth]{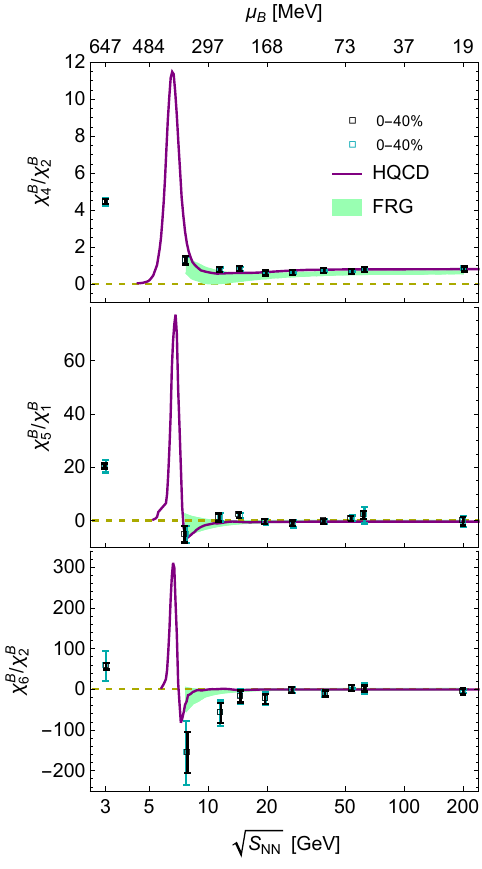}
  \caption{Comparison between the baryon number susceptibilities along two fitted chemical freeze-out lines and the STAR data of net-proton distributions for centrality $0-5\%$ (left)~\cite{Xu:2016mqs,STAR:2020tga,STAR:2021fge,Ko:2023whx} and $0-40\%$ (right)~\cite{STAR:2022vlo} Au-Au collisions, respectively. The error bars in the left panel (grey and brown) and the right panel (black and cyan) respectively denote the statistical and systematic uncertainties. The FRG results~\cite{Fu:2021oaw} are shown in green bands.}\label{fig6}
\end{figure}

In Fig.~\ref{fig6}, we present a direct comparison between the baryon number susceptibilities computed using our fitted chemical freeze-out lines and the experimental data from STAR with centrality ranges of $0-5\%$  (left)\cite{Xu:2016mqs,STAR:2020tga,STAR:2021fge} and $0-40\%$ (right)\cite{STAR:2022vlo}. The different error bars of Fig.~\ref{fig6}, grey and brown in the left panel as well as black and cyan in the right panel, corresponding to the statistical and systematic uncertainties of the experimental data points, respectively. Moreover, we also include the FRG results denoted as green bands~\cite{Fu:2021oaw}. Our results demonstrate quantitative agreement with both STAR and FRG results for second and third-order baryon number susceptibilities in the collision energy range of $\sqrt{S_{NN}}\approx 12-200~\text{GeV}$. Interestingly, at lower collision energies ($\sqrt{S_{NN}}\approx 5-10~\text{GeV}$), our results reveal a peak structure for both $\chi_2^B/\chi_1^B$ and $\chi_3^B/\chi_2^B$, which is not reflected in the experimental data. From fourth to sixth order, our results show quantitative agreement with experimental data and FRG results in the collision energy range of $\sqrt{S_{NN}}\approx 7.7-200~\text{GeV}$ for both centralities. Remarkably, the ratios $\chi_m^B/\chi_n^B$ with $m>n$ in our model form a peak structure around $\sqrt{S_{NN}}\approx 5-10~\text{GeV}$, with the peak becoming sharper and larger as we progress to higher orders. 

We also find that at collision energies below $\sqrt{S_{NN}}\approx 5~\text{GeV}$, the ratios $\chi_m^B/\chi_n^B$ obtained from our model approach zero, which deviates from the STAR data of centrality $0-40\%$~\cite{STAR:2022vlo} (see the right column of Fig.~\ref{fig6}). Such discrepancy observed at low collision energies could stem from several factors, including non-equilibrium effects of low energy collisions and complex experimental environments (such as rotation and magnetic field\,\footnote{The rotation and magnetic field effects in holographic QCD was investigated \emph{e.g.} in~\cite{Rougemont:2015oea, Finazzo:2016mhm, Critelli:2016fvr, Chen:2020ath, Fang:2021ucy, Zhao:2022uxc}.} in non-centric collisions). Therefore, further studies are necessary to investigate the role of these effects on the above-observed discrepancy. 
\begin{figure}[htbp]
\centering
\includegraphics[width=.47\textwidth]{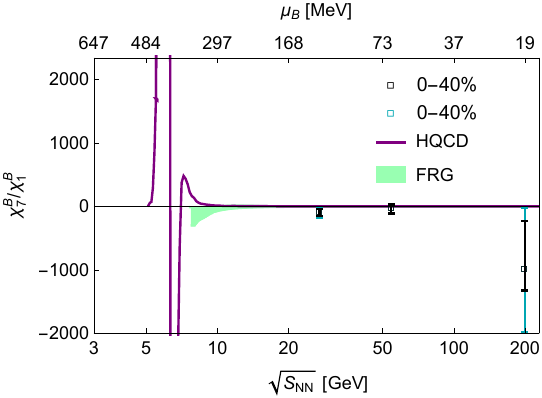}
\qquad
\includegraphics[width=.47\textwidth]{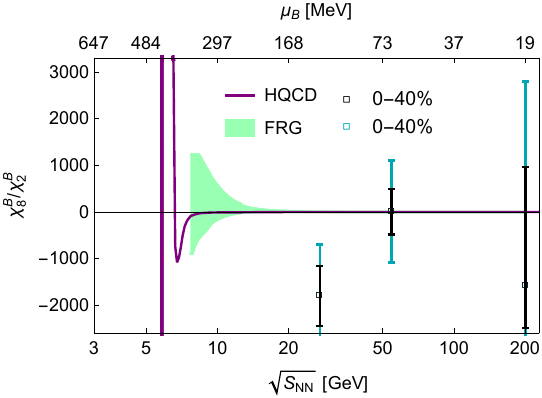}
  \caption{Comparison between the baryon number susceptibilities $\chi_7^B/\chi_1^B$ and $\chi_8^B/\chi_2^B$ along fitted chemical freeze-out line ``fit-2" and the STAR data of net-proton distributions in $0-40\%$~\cite{Pandav:2023lis} centrality Au-Au collisions. The black and cyan error bars correspond to the statistical and systematic uncertainties of the experimental data points, respectively.  The FRG results~\cite{Fu:2021oaw} are denoted by green bands.}\label{fig7}
\end{figure}

Thus far, there are few experimental data for higher-order susceptibilities. Our prediction for $\chi_7^B/\chi_1^B$ and $\chi_8^B/\chi_2^B$ is presented in Fig.~\ref{fig7}. The experimental data from STAR with the centrality of $0-40\%$~\cite{Pandav:2023lis} and the FRG results~\cite{Fu:2021oaw} are also included as a comparison. Current STAR data only includes points with collision energies of $27$, $54.4$, and $200~\text{GeV}$, and the uncertainty is relatively high, making it challenging to identify non-monotonic behavior from the data. Nevertheless, our results suggest that higher-order baryon number susceptibilities will exhibit more pronounced non-monotonic behavior, with additional peaks and dips appearing along the freeze-out line. These features could be potentially observed in future experiments with improved precision.

%%%%%%%%%%%%%%%%%%%%%%%%%%%%%%%%%%%%%%%%%%%%%%
\section{Conclusion}\label{sec:05}
%%%%%%%%%%%%%%%%%%%%%%%%%%%%%%%%%%%%%%%%%%%%%%

In this study, we have investigated the behavior of higher-order baryon number susceptibilities ($\chi_n^B$) at finite temperature and baryon chemical potential using a quantitative holographic QCD model that has been calibrated with lattice QCD data. We have observed a quantitative agreement between our results and the HotQCD lattice data at $\mu_B=0$, see Figs.~\ref{fig1} and~\ref{fig2}. The examination of $\chi_{10}^B$ and $\chi_{12}^B$ will be possible in the future once relevant lattice data becomes available.

To investigate the critical physics associated with the CEP, we have analyzed the dependence of various ratios of $\chi_n^B$ on collision energy along the chemical freeze-out line. Our findings demonstrate a quantitative agreement with experimental data from STAR net-proton high moments in $0-5\%$ centrality and $0-40\%$ centrality Au-Au collisions, as well as FRG results for second and third-order baryon number susceptibilities within the range $\sqrt{S_{NN}}\approx 12-200~\text{GeV}$. At lower collision energies, we have observed a clear peak structure from $\chi_2^B/\chi_1^B$ and $\chi_3^B/\chi_2^B$ around $\sqrt{S_{NN}}\approx 5-10~\text{GeV}$ (see Fig.~\ref{fig6}), while experimental data is accumulating to check this feature. Regarding the fourth to sixth-order ratios, our results have exhibited quantitative agreement with experimental data and FRG results over a broad collision energy range of $\sqrt{S_{NN}}\approx 7.7-200~\text{GeV}$. We have found that all ratios $\chi_m^B/\chi_n^B$ with $m>n$ display peak structures around $\sqrt{S_{NN}}\approx 5-10~\text{GeV}$, characterized by sharper and higher peaks as $m$ is increased.

However, for low collision energies ($\sqrt{S_{NN}}<5~\text{GeV}$), the ratios $\chi_m^B/\chi_n^B$ tend to approach zero, which deviates from the STAR data. This discrepancy can potentially be attributed to non-equilibrium effects present in low-energy collisions, as well as the influence of complex experimental environments, such as rotation and magnetic field effects in non-central collisions. Notably, it has been proposed that rotation might have a significant impact on the results obtained from low-energy collisions. Therefore, we suggest that future experiments with measurements conducted in the low collision energy range of $\sqrt{S_{NN}}\approx 1-10~\text{GeV}$, along with reduced experimental uncertainties, will unveil further non-monotonic behavioral signals that could aid in the precise determination of the location of the CEP.

%\appendix
%\section{Some title}

%%%%%%%%%%%%%%%%%%%%%%%%%%%%%%%%%%%%%%%%%%%%%%
\section{Acknowledgments}\label{sec:06}
%%%%%%%%%%%%%%%%%%%%%%%%%%%%%%%%%%%%%%%%%%%%%%

We would like to thank Yong Cai, Wei-Jie Fu, De-Fu Hou, Cheng-Ming Li, Shu Lin, Peng Liu, Xiaofeng Luo, Shi Pu, and Yi-Bo Yang for their helpful discussion. S.H. would appreciate the financial support from the Fundamental Research Funds for the Central Universities and Max Planck Partner Group, as well as the Natural Science Foundation of China (NSFC) Grants No.~12075101 and No.~12235016. L.L. was partially supported by the NSFC Grants No.12122513, No.12075298, and No.12047503, and by the Chinese Academy of Sciences Project for Young Scientists in Basic Research YSBR-006.

%\bibliographystyle{utphys}
%\bibliography{ref.bib}

\begin{thebibliography}{10}
	
	\bibitem{Borsanyi:2010cj}
	S.~Borsanyi, G.~Endrodi, Z.~Fodor, A.~Jakovac, S.~D. Katz, S.~Krieg, C.~Ratti,
	and K.~K. Szabo, ``{The QCD equation of state with dynamical quarks},''
	\href{http://dx.doi.org/10.1007/JHEP11(2010)077}{{\em JHEP} {\bfseries 11}
		(2010) 077}, \href{http://arxiv.org/abs/1007.2580}{{\ttfamily arXiv:1007.2580
			[hep-lat]}}.
	
	\bibitem{Borsanyi:2013bia}
	S.~Borsanyi, Z.~Fodor, C.~Hoelbling, S.~D. Katz, S.~Krieg, and K.~K. Szabo,
	``{Full result for the QCD equation of state with 2+1 flavors},''
	\href{http://dx.doi.org/10.1016/j.physletb.2014.01.007}{{\em Phys. Lett. B}
		{\bfseries 730} (2014) 99--104},
	\href{http://arxiv.org/abs/1309.5258}{{\ttfamily arXiv:1309.5258 [hep-lat]}}.
	
	\bibitem{HotQCD:2014kol}
	{\bfseries HotQCD} Collaboration, A.~Bazavov {\em et~al.}, ``{Equation of state
		in ( 2+1 )-flavor QCD},''
	\href{http://dx.doi.org/10.1103/PhysRevD.90.094503}{{\em Phys. Rev. D}
		{\bfseries 90} (2014) 094503},
	\href{http://arxiv.org/abs/1407.6387}{{\ttfamily arXiv:1407.6387 [hep-lat]}}.
	
	\bibitem{Xin:2014ela}
	X.-y. Xin, S.-x. Qin, and Y.-x. Liu, ``{Quark number fluctuations at finite
		temperature and finite chemical potential via the Dyson-Schwinger equation
		approach},'' \href{http://dx.doi.org/10.1103/PhysRevD.90.076006}{{\em Phys.
			Rev. D} {\bfseries 90} no.~7, (2014) 076006},
	\href{http://arxiv.org/abs/2109.09935}{{\ttfamily arXiv:2109.09935
			[hep-ph]}}.
	
	\bibitem{Gao:2016qkh}
	F.~Gao and Y.-x. Liu, ``{QCD phase transitions via a refined truncation of
		Dyson-Schwinger equations},''
	\href{http://dx.doi.org/10.1103/PhysRevD.94.076009}{{\em Phys. Rev. D}
		{\bfseries 94} no.~7, (2016) 076009},
	\href{http://arxiv.org/abs/1607.01675}{{\ttfamily arXiv:1607.01675
			[hep-ph]}}.
	
	\bibitem{Qin:2010nq}
	S.-x. Qin, L.~Chang, H.~Chen, Y.-x. Liu, and C.~D. Roberts, ``{Phase diagram
		and critical endpoint for strongly-interacting quarks},''
	\href{http://dx.doi.org/10.1103/PhysRevLett.106.172301}{{\em Phys. Rev.
			Lett.} {\bfseries 106} (2011) 172301},
	\href{http://arxiv.org/abs/1011.2876}{{\ttfamily arXiv:1011.2876 [nucl-th]}}.
	
	\bibitem{Shi:2014zpa}
	C.~Shi, Y.-L. Wang, Y.~Jiang, Z.-F. Cui, and H.-S. Zong, ``{Locate QCD Critical
		End Point in a Continuum Model Study},''
	\href{http://dx.doi.org/10.1007/JHEP07(2014)014}{{\em JHEP} {\bfseries 07}
		(2014) 014}, \href{http://arxiv.org/abs/1403.3797}{{\ttfamily arXiv:1403.3797
			[hep-ph]}}.
	
	\bibitem{Fischer:2014ata}
	C.~S. Fischer, J.~Luecker, and C.~A. Welzbacher, ``{Phase structure of three
		and four flavor QCD},''
	\href{http://dx.doi.org/10.1103/PhysRevD.90.034022}{{\em Phys. Rev. D}
		{\bfseries 90} no.~3, (2014) 034022},
	\href{http://arxiv.org/abs/1405.4762}{{\ttfamily arXiv:1405.4762 [hep-ph]}}.
	
	\bibitem{Gao:2020qsj}
	F.~Gao and J.~M. Pawlowski, ``{QCD phase structure from functional methods},''
	\href{http://dx.doi.org/10.1103/PhysRevD.102.034027}{{\em Phys. Rev. D}
		{\bfseries 102} no.~3, (2020) 034027},
	\href{http://arxiv.org/abs/2002.07500}{{\ttfamily arXiv:2002.07500
			[hep-ph]}}.
	
	\bibitem{Asakawa:1989bq}
	M.~Asakawa and K.~Yazaki, ``{Chiral Restoration at Finite Density and
		Temperature},'' \href{http://dx.doi.org/10.1016/0375-9474(89)90002-X}{{\em
			Nucl. Phys. A} {\bfseries 504} (1989) 668--684}.
	
	\bibitem{Schwarz:1999dj}
	T.~M. Schwarz, S.~P. Klevansky, and G.~Papp, ``{The Phase diagram and bulk
		thermodynamical quantities in the NJL model at finite temperature and
		density},'' \href{http://dx.doi.org/10.1103/PhysRevC.60.055205}{{\em Phys.
			Rev. C} {\bfseries 60} (1999) 055205},
	\href{http://arxiv.org/abs/nucl-th/9903048}{{\ttfamily
			arXiv:nucl-th/9903048}}.
	
	\bibitem{Li:2018ygx}
	Z.~Li, K.~Xu, X.~Wang, and M.~Huang, ``{The kurtosis of net baryon number
		fluctuations from a realistic
		Polyakov\textendash{}Nambu\textendash{}Jona-Lasinio model along the
		experimental freeze-out line},''
	\href{http://dx.doi.org/10.1140/epjc/s10052-019-6703-x}{{\em Eur. Phys. J. C}
		{\bfseries 79} no.~3, (2019) 245},
	\href{http://arxiv.org/abs/1801.09215}{{\ttfamily arXiv:1801.09215
			[hep-ph]}}.
	
	\bibitem{Zhuang:2000ub}
	P.~Zhuang, M.~Huang, and Z.~Yang, ``{Density effect on hadronization of a quark
		plasma},'' \href{http://dx.doi.org/10.1103/PhysRevC.62.054901}{{\em Phys.
			Rev. C} {\bfseries 62} (2000) 054901},
	\href{http://arxiv.org/abs/nucl-th/0008043}{{\ttfamily
			arXiv:nucl-th/0008043}}.
	
	\bibitem{Fu:2019hdw}
	W.-j. Fu, J.~M. Pawlowski, and F.~Rennecke, ``{QCD phase structure at finite
		temperature and density},''
	\href{http://dx.doi.org/10.1103/PhysRevD.101.054032}{{\em Phys. Rev. D}
		{\bfseries 101} no.~5, (2020) 054032},
	\href{http://arxiv.org/abs/1909.02991}{{\ttfamily arXiv:1909.02991
			[hep-ph]}}.
	
	\bibitem{Zhang:2017icm}
	H.~Zhang, D.~Hou, T.~Kojo, and B.~Qin, ``{Functional renormalization group
		study of the quark-meson model with $\omega$ meson},''
	\href{http://dx.doi.org/10.1103/PhysRevD.96.114029}{{\em Phys. Rev. D}
		{\bfseries 96} no.~11, (2017) 114029},
	\href{http://arxiv.org/abs/1709.05654}{{\ttfamily arXiv:1709.05654
			[hep-ph]}}.
	
	\bibitem{Fu:2021oaw}
	W.-j. Fu, X.~Luo, J.~M. Pawlowski, F.~Rennecke, R.~Wen, and S.~Yin,
	``{Hyper-order baryon number fluctuations at finite temperature and
		density},'' \href{http://dx.doi.org/10.1103/PhysRevD.104.094047}{{\em Phys.
			Rev. D} {\bfseries 104} no.~9, (2021) 094047},
	\href{http://arxiv.org/abs/2101.06035}{{\ttfamily arXiv:2101.06035
			[hep-ph]}}.
	
	\bibitem{Vovchenko:2017gkg}
	V.~Vovchenko, J.~Steinheimer, O.~Philipsen, and H.~Stoecker, ``{Cluster
		Expansion Model for QCD Baryon Number Fluctuations: No Phase Transition at
		$\mu_B / T < \pi$},''
	\href{http://dx.doi.org/10.1103/PhysRevD.97.114030}{{\em Phys. Rev. D}
		{\bfseries 97} no.~11, (2018) 114030},
	\href{http://arxiv.org/abs/1711.01261}{{\ttfamily arXiv:1711.01261
			[hep-ph]}}.
	
	\bibitem{Borsanyi:2020fev}
	S.~Borsanyi, Z.~Fodor, J.~N. Guenther, R.~Kara, S.~D. Katz, P.~Parotto,
	A.~Pasztor, C.~Ratti, and K.~K. Szabo, ``{QCD Crossover at Finite Chemical
		Potential from Lattice Simulations},''
	\href{http://dx.doi.org/10.1103/PhysRevLett.125.052001}{{\em Phys. Rev.
			Lett.} {\bfseries 125} no.~5, (2020) 052001},
	\href{http://arxiv.org/abs/2002.02821}{{\ttfamily arXiv:2002.02821
			[hep-lat]}}.
	
	\bibitem{Bazavov:2020bjn}
	A.~Bazavov {\em et~al.}, ``{Skewness, kurtosis, and the fifth and sixth order
		cumulants of net baryon-number distributions from lattice QCD confront
		high-statistics STAR data},''
	\href{http://dx.doi.org/10.1103/PhysRevD.101.074502}{{\em Phys. Rev. D}
		{\bfseries 101} no.~7, (2020) 074502},
	\href{http://arxiv.org/abs/2001.08530}{{\ttfamily arXiv:2001.08530
			[hep-lat]}}.
	
	\bibitem{Borsanyi:2021sxv}
	S.~Bors\'anyi, Z.~Fodor, J.~N. Guenther, R.~Kara, S.~D. Katz, P.~Parotto,
	A.~P\'asztor, C.~Ratti, and K.~K. Szab\'o, ``{Lattice QCD equation of state
		at finite chemical potential from an alternative expansion scheme},''
	\href{http://dx.doi.org/10.1103/PhysRevLett.126.232001}{{\em Phys. Rev.
			Lett.} {\bfseries 126} no.~23, (2021) 232001},
	\href{http://arxiv.org/abs/2102.06660}{{\ttfamily arXiv:2102.06660
			[hep-lat]}}.
	
	\bibitem{Bollweg:2022fqq}
	D.~Bollweg, D.~A. Clarke, J.~Goswami, O.~Kaczmarek, F.~Karsch, S.~Mukherjee,
	P.~Petreczky, C.~Schmidt, and S.~Sharma, ``{Equation of state and speed of
		sound of (2+1)-flavor QCD in strangeness-neutral matter at non-vanishing net
		baryon-number density},'' \href{http://arxiv.org/abs/2212.09043}{{\ttfamily
			arXiv:2212.09043 [hep-lat]}}.
	
	\bibitem{Philipsen:2021qji}
	O.~Philipsen, ``{Lattice Constraints on the QCD Chiral Phase Transition at
		Finite Temperature and Baryon Density},''
	\href{http://dx.doi.org/10.3390/sym13112079}{{\em Symmetry} {\bfseries 13}
		no.~11, (2021) 2079}, \href{http://arxiv.org/abs/2111.03590}{{\ttfamily
			arXiv:2111.03590 [hep-lat]}}.
	
	\bibitem{Schaefer:2006ds}
	B.-J. Schaefer and J.~Wambach, ``{Susceptibilities near the QCD (tri)critical
		point},'' \href{http://dx.doi.org/10.1103/PhysRevD.75.085015}{{\em Phys. Rev.
			D} {\bfseries 75} (2007) 085015},
	\href{http://arxiv.org/abs/hep-ph/0603256}{{\ttfamily arXiv:hep-ph/0603256}}.
	
	\bibitem{Asakawa:2009aj}
	M.~Asakawa, S.~Ejiri, and M.~Kitazawa, ``{Third moments of conserved charges as
		probes of QCD phase structure},''
	\href{http://dx.doi.org/10.1103/PhysRevLett.103.262301}{{\em Phys. Rev.
			Lett.} {\bfseries 103} (2009) 262301},
	\href{http://arxiv.org/abs/0904.2089}{{\ttfamily arXiv:0904.2089 [nucl-th]}}.
	
	\bibitem{Schaefer:2011ex}
	B.~J. Schaefer and M.~Wagner, ``{QCD critical region and higher moments for
		three flavor models},''
	\href{http://dx.doi.org/10.1103/PhysRevD.85.034027}{{\em Phys. Rev. D}
		{\bfseries 85} (2012) 034027},
	\href{http://arxiv.org/abs/1111.6871}{{\ttfamily arXiv:1111.6871 [hep-ph]}}.
	
	\bibitem{Fan:2016ovc}
	W.~Fan, X.~Luo, and H.-S. Zong, ``{Mapping the QCD phase diagram with
		susceptibilities of conserved charges within Nambu\textendash{}Jona-Lasinio
		model},'' \href{http://dx.doi.org/10.1142/S0217751X17500610}{{\em Int. J.
			Mod. Phys. A} {\bfseries 32} no.~11, (2017) 1750061},
	\href{http://arxiv.org/abs/1608.07903}{{\ttfamily arXiv:1608.07903
			[hep-ph]}}.
	
	\bibitem{Portillo:2016fso}
	I.~Portillo, ``{Susceptibilities from a black hole engineered EoS with a
		critical point},''
	\href{http://dx.doi.org/10.1088/1742-6596/832/1/012041}{{\em J. Phys. Conf.
			Ser.} {\bfseries 832} no.~1, (2017) 012041},
	\href{http://arxiv.org/abs/1610.09981}{{\ttfamily arXiv:1610.09981
			[nucl-th]}}.
	
	\bibitem{Fan:2017mrk}
	W.~Fan, X.~Luo, and H.~Zong, ``{Probing the QCD phase structure with higher
		order baryon number susceptibilities within the NJL model},''
	\href{http://dx.doi.org/10.1088/1674-1137/43/3/033103}{{\em Chin. Phys. C}
		{\bfseries 43} no.~3, (2019) 033103},
	\href{http://arxiv.org/abs/1702.08674}{{\ttfamily arXiv:1702.08674
			[hep-ph]}}.
	
	\bibitem{Li:2017ple}
	Z.~Li, Y.~Chen, D.~Li, and M.~Huang, ``{Locating the QCD critical end point
		through the peaked baryon number susceptibilities along the freeze-out
		line},'' \href{http://dx.doi.org/10.1088/1674-1137/42/1/013103}{{\em Chin.
			Phys. C} {\bfseries 42} no.~1, (2018) 013103},
	\href{http://arxiv.org/abs/1706.02238}{{\ttfamily arXiv:1706.02238
			[hep-ph]}}.
	
	\bibitem{Zhao:2023xpj}
	Y.-P. Zhao, C.-Y. Wang, S.-Y. Zuo, and C.-M. Li, ``{Nonextensive effects on QCD
		chiral phase diagram and baryon-number fluctuations within
		Polyakov-Nambu-Jona-Lasinio model*},''
	\href{http://dx.doi.org/10.1088/1674-1137/acbf2a}{{\em Chin. Phys. C}
		{\bfseries 47} no.~5, (2023) 053103},
	\href{http://arxiv.org/abs/2302.12010}{{\ttfamily arXiv:2302.12010
			[hep-ph]}}.
	
	\bibitem{Stephanov:2011pb}
	M.~A. Stephanov, ``{On the sign of kurtosis near the QCD critical point},''
	\href{http://dx.doi.org/10.1103/PhysRevLett.107.052301}{{\em Phys. Rev.
			Lett.} {\bfseries 107} (2011) 052301},
	\href{http://arxiv.org/abs/1104.1627}{{\ttfamily arXiv:1104.1627 [hep-ph]}}.
	
	\bibitem{STAR:2017sal}
	{\bfseries STAR} Collaboration, L.~Adamczyk {\em et~al.}, ``{Bulk Properties of
		the Medium Produced in Relativistic Heavy-Ion Collisions from the Beam Energy
		Scan Program},'' \href{http://dx.doi.org/10.1103/PhysRevC.96.044904}{{\em
			Phys. Rev. C} {\bfseries 96} no.~4, (2017) 044904},
	\href{http://arxiv.org/abs/1701.07065}{{\ttfamily arXiv:1701.07065
			[nucl-ex]}}.
	
	\bibitem{Portillo:2017gfk}
	I.~Portillo, ``{Baryon susceptibilities from a holographic equation of
		state},'' \href{http://dx.doi.org/10.1016/j.nuclphysa.2017.05.012}{{\em Nucl.
			Phys. A} {\bfseries 967} (2017) 916--919},
	\href{http://arxiv.org/abs/1705.01021}{{\ttfamily arXiv:1705.01021
			[nucl-th]}}.
	
	\bibitem{Wang:2018sur}
	X.~Wang, M.~Wei, Z.~Li, and M.~Huang, ``{Quark matter under rotation in the NJL
		model with vector interaction},''
	\href{http://dx.doi.org/10.1103/PhysRevD.99.016018}{{\em Phys. Rev. D}
		{\bfseries 99} no.~1, (2019) 016018},
	\href{http://arxiv.org/abs/1808.01931}{{\ttfamily arXiv:1808.01931
			[hep-ph]}}.
	
	\bibitem{Huang:2023ogw}
	M.~Huang and P.~Zhuang, ``{QCD Matter and Phase Transitions under Extreme
		Conditions},'' \href{http://dx.doi.org/10.3390/sym15020541}{{\em Symmetry}
		{\bfseries 15} no.~2, (2023) 541}.
	
	\bibitem{Luo:2017faz}
	X.~Luo and N.~Xu, ``{Search for the QCD Critical Point with Fluctuations of
		Conserved Quantities in Relativistic Heavy-Ion Collisions at RHIC : An
		Overview},'' \href{http://dx.doi.org/10.1007/s41365-017-0257-0}{{\em Nucl.
			Sci. Tech.} {\bfseries 28} no.~8, (2017) 112},
	\href{http://arxiv.org/abs/1701.02105}{{\ttfamily arXiv:1701.02105
			[nucl-ex]}}.
	
	\bibitem{STAR:2021iop}
	{\bfseries STAR} Collaboration, M.~Abdallah {\em et~al.}, ``{Cumulants and
		correlation functions of net-proton, proton, and antiproton multiplicity
		distributions in Au+Au collisions at energies available at the BNL
		Relativistic Heavy Ion Collider},''
	\href{http://dx.doi.org/10.1103/PhysRevC.104.024902}{{\em Phys. Rev. C}
		{\bfseries 104} no.~2, (2021) 024902},
	\href{http://arxiv.org/abs/2101.12413}{{\ttfamily arXiv:2101.12413
			[nucl-ex]}}.
	
	\bibitem{STAR:2020tga}
	{\bfseries STAR} Collaboration, J.~Adam {\em et~al.}, ``{Nonmonotonic Energy
		Dependence of Net-Proton Number Fluctuations},''
	\href{http://dx.doi.org/10.1103/PhysRevLett.126.092301}{{\em Phys. Rev.
			Lett.} {\bfseries 126} no.~9, (2021) 092301},
	\href{http://arxiv.org/abs/2001.02852}{{\ttfamily arXiv:2001.02852
			[nucl-ex]}}.
	
	\bibitem{STAR:2021fge}
	{\bfseries STAR} Collaboration, M.~S. Abdallah {\em et~al.}, ``{Measurements of
		Proton High Order Cumulants in $\sqrt{s_{_{\mathrm{NN}}}}$ = 3 GeV Au+Au
		Collisions and Implications for the QCD Critical Point},''
	\href{http://dx.doi.org/10.1103/PhysRevLett.128.202303}{{\em Phys. Rev.
			Lett.} {\bfseries 128} no.~20, (2022) 202303},
	\href{http://arxiv.org/abs/2112.00240}{{\ttfamily arXiv:2112.00240
			[nucl-ex]}}.
	
	\bibitem{Pandav:2022xxx}
	A.~Pandav, D.~Mallick, and B.~Mohanty, ``{Search for the QCD critical point in
		high energy nuclear collisions},''
	\href{http://dx.doi.org/10.1016/j.ppnp.2022.103960}{{\em Prog. Part. Nucl.
			Phys.} {\bfseries 125} (2022) 103960},
	\href{http://arxiv.org/abs/2203.07817}{{\ttfamily arXiv:2203.07817
			[nucl-ex]}}.
	
	\bibitem{Zhang:2022evi}
	{\bfseries STAR} Collaboration, Y.~Zhang, ``{Higher-order Proton Cumulants in
		Au+Au Collisions at \(\sqrt {s_{NN}} = 3\) GeV from RHIC-STAR},''
	\href{http://dx.doi.org/10.5506/APhysPolBSupp.16.1-A45}{{\em Acta Phys.
			Polon. Supp.} {\bfseries 16} no.~1, (2023) 45},
	\href{http://arxiv.org/abs/2208.05686}{{\ttfamily arXiv:2208.05686
			[nucl-ex]}}.
	
	\bibitem{STAR:2022vlo}
	{\bfseries STAR} Collaboration, B.~Aboona {\em et~al.}, ``{Beam Energy
		Dependence of Fifth and Sixth-Order Net-proton Number Fluctuations in Au+Au
		Collisions at RHIC},''
	\href{http://dx.doi.org/10.1103/PhysRevLett.130.082301}{{\em Phys. Rev.
			Lett.} {\bfseries 130} no.~8, (2023) 082301},
	\href{http://arxiv.org/abs/2207.09837}{{\ttfamily arXiv:2207.09837
			[nucl-ex]}}.
	
	\bibitem{Pandav:2023lis}
	{\bfseries STAR} Collaboration, A.~Pandav, ``{Seventh and eighth-order
		cumulants of net-proton multiplicity distributions in heavy-ion collisions at
		RHIC-STAR},'' \href{http://dx.doi.org/10.1051/epjconf/202327601006}{{\em EPJ
			Web Conf.} {\bfseries 276} (2023) 01006}.
	
	\bibitem{Cai:2022omk}
	R.-G. Cai, S.~He, L.~Li, and Y.-X. Wang, ``{Probing QCD critical point and
		induced gravitational wave by black hole physics},''
	\href{http://dx.doi.org/10.1103/PhysRevD.106.L121902}{{\em Phys. Rev. D}
		{\bfseries 106} no.~12, (2022) L121902},
	\href{http://arxiv.org/abs/2201.02004}{{\ttfamily arXiv:2201.02004
			[hep-th]}}.
	
	\bibitem{Rougemont:2015ona}
	R.~Rougemont, J.~Noronha, and J.~Noronha-Hostler, ``{Suppression of baryon
		diffusion and transport in a baryon rich strongly coupled quark-gluon
		plasma},'' \href{http://dx.doi.org/10.1103/PhysRevLett.115.202301}{{\em Phys.
			Rev. Lett.} {\bfseries 115} no.~20, (2015) 202301},
	\href{http://arxiv.org/abs/1507.06972}{{\ttfamily arXiv:1507.06972
			[hep-ph]}}.
	
	\bibitem{Critelli:2017oub}
	R.~Critelli, J.~Noronha, J.~Noronha-Hostler, I.~Portillo, C.~Ratti, and
	R.~Rougemont, ``{Critical point in the phase diagram of primordial
		quark-gluon matter from black hole physics},''
	\href{http://dx.doi.org/10.1103/PhysRevD.96.096026}{{\em Phys. Rev. D}
		{\bfseries 96} no.~9, (2017) 096026},
	\href{http://arxiv.org/abs/1706.00455}{{\ttfamily arXiv:1706.00455
			[nucl-th]}}.
	
	\bibitem{DeWolfe:2010he}
	O.~DeWolfe, S.~S. Gubser, and C.~Rosen, ``{A holographic critical point},''
	\href{http://dx.doi.org/10.1103/PhysRevD.83.086005}{{\em Phys. Rev. D}
		{\bfseries 83} (2011) 086005},
	\href{http://arxiv.org/abs/1012.1864}{{\ttfamily arXiv:1012.1864 [hep-th]}}.
	
	\bibitem{DeWolfe:2011ts}
	O.~DeWolfe, S.~S. Gubser, and C.~Rosen, ``{Dynamic critical phenomena at a
		holographic critical point},''
	\href{http://dx.doi.org/10.1103/PhysRevD.84.126014}{{\em Phys. Rev. D}
		{\bfseries 84} (2011) 126014},
	\href{http://arxiv.org/abs/1108.2029}{{\ttfamily arXiv:1108.2029 [hep-th]}}.
	
	\bibitem{Cai:2012xh}
	R.-G. Cai, S.~He, and D.~Li, ``{A hQCD model and its phase diagram in
		Einstein-Maxwell-Dilaton system},''
	\href{http://dx.doi.org/10.1007/JHEP03(2012)033}{{\em JHEP} {\bfseries 03}
		(2012) 033}, \href{http://arxiv.org/abs/1201.0820}{{\ttfamily arXiv:1201.0820
			[hep-th]}}.
	
	\bibitem{Cai:2012eh}
	R.-G. Cai, S.~Chakrabortty, S.~He, and L.~Li, ``{Some aspects of QGP phase in a
		hQCD model},'' \href{http://dx.doi.org/10.1007/JHEP02(2013)068}{{\em JHEP}
		{\bfseries 02} (2013) 068}, \href{http://arxiv.org/abs/1209.4512}{{\ttfamily
			arXiv:1209.4512 [hep-th]}}.
	
	\bibitem{Finazzo:2013efa}
	S.~I. Finazzo and J.~Noronha, ``{Holographic calculation of the electric
		conductivity of the strongly coupled quark-gluon plasma near the
		deconfinement transition},''
	\href{http://dx.doi.org/10.1103/PhysRevD.89.106008}{{\em Phys. Rev. D}
		{\bfseries 89} no.~10, (2014) 106008},
	\href{http://arxiv.org/abs/1311.6675}{{\ttfamily arXiv:1311.6675 [hep-th]}}.
	
	\bibitem{Yang:2014bqa}
	Y.~Yang and P.-H. Yuan, ``{A Refined Holographic QCD Model and QCD Phase
		Structure},'' \href{http://dx.doi.org/10.1007/JHEP11(2014)149}{{\em JHEP}
		{\bfseries 11} (2014) 149}, \href{http://arxiv.org/abs/1406.1865}{{\ttfamily
			arXiv:1406.1865 [hep-th]}}.
	
	\bibitem{Chen:2017cyc}
	Y.~Chen, M.~Huang, and Q.-S. Yan, ``{Gravitation waves from QCD and electroweak
		phase transitions},'' \href{http://dx.doi.org/10.1007/JHEP05(2018)178}{{\em
			JHEP} {\bfseries 05} (2018) 178},
	\href{http://arxiv.org/abs/1712.03470}{{\ttfamily arXiv:1712.03470
			[hep-ph]}}.
	
	\bibitem{Knaute:2017opk}
	J.~Knaute, R.~Yaresko, and B.~K\"ampfer, ``{Holographic QCD phase diagram with
		critical point from Einstein\textendash{}Maxwell-dilaton dynamics},''
	\href{http://dx.doi.org/10.1016/j.physletb.2018.01.053}{{\em Phys. Lett. B}
		{\bfseries 778} (2018) 419--425},
	\href{http://arxiv.org/abs/1702.06731}{{\ttfamily arXiv:1702.06731
			[hep-ph]}}.
	
	\bibitem{Fang:2018axm}
	Z.~Fang, Y.-L. Wu, and L.~Zhang, ``{Chiral phase transition and QCD phase
		diagram from AdS/QCD},''
	\href{http://dx.doi.org/10.1103/PhysRevD.99.034028}{{\em Phys. Rev. D}
		{\bfseries 99} no.~3, (2019) 034028},
	\href{http://arxiv.org/abs/1810.12525}{{\ttfamily arXiv:1810.12525
			[hep-ph]}}.
	
	\bibitem{Ballon-Bayona:2020xls}
	A.~Ballon-Bayona, H.~Boschi-Filho, E.~F. Capossoli, and D.~M. Rodrigues,
	``{Criticality from Einstein-Maxwell-dilaton holography at finite temperature
		and density},'' \href{http://dx.doi.org/10.1103/PhysRevD.102.126003}{{\em
			Phys. Rev. D} {\bfseries 102} no.~12, (2020) 126003},
	\href{http://arxiv.org/abs/2006.08810}{{\ttfamily arXiv:2006.08810
			[hep-th]}}.
	
	\bibitem{Li:2020hau}
	M.-W. Li, Y.~Yang, and P.-H. Yuan, ``{Analytic Study on Chiral Phase Transition
		in Holographic QCD},'' \href{http://dx.doi.org/10.1007/JHEP02(2021)055}{{\em
			JHEP} {\bfseries 02} (2021) 055},
	\href{http://arxiv.org/abs/2009.05694}{{\ttfamily arXiv:2009.05694
			[hep-th]}}.
	
	\bibitem{Grefa:2021qvt}
	J.~Grefa, J.~Noronha, J.~Noronha-Hostler, I.~Portillo, C.~Ratti, and
	R.~Rougemont, ``{Hot and dense quark-gluon plasma thermodynamics from
		holographic black holes},''
	\href{http://dx.doi.org/10.1103/PhysRevD.104.034002}{{\em Phys. Rev. D}
		{\bfseries 104} no.~3, (2021) 034002},
	\href{http://arxiv.org/abs/2102.12042}{{\ttfamily arXiv:2102.12042
			[nucl-th]}}.
	
	\bibitem{He:2022amv}
	S.~He, L.~Li, Z.~Li, and S.-J. Wang, ``{Gravitational Waves and Primordial
		Black Hole Productions from Gluodynamics},''
	\href{http://arxiv.org/abs/2210.14094}{{\ttfamily arXiv:2210.14094
			[hep-ph]}}.
	
	\bibitem{Grefa:2022fpu}
	J.~Grefa, M.~Hippert, J.~Noronha, J.~Noronha-Hostler, I.~Portillo, C.~Ratti,
	and R.~Rougemont, ``{QCD Equilibrium and Dynamical Properties from
		Holographic Black Holes},''
	\href{http://dx.doi.org/10.31349/SuplRevMexFis.3.040910}{{\em Rev. Mex. Fis.
			Suppl.} {\bfseries 3} no.~4, (2022) 040910},
	\href{http://arxiv.org/abs/2207.12564}{{\ttfamily arXiv:2207.12564
			[nucl-th]}}.
	
	\bibitem{HotQCD:2012fhj}
	{\bfseries HotQCD} Collaboration, A.~Bazavov {\em et~al.}, ``{Fluctuations and
		Correlations of net baryon number, electric charge, and strangeness: A
		comparison of lattice QCD results with the hadron resonance gas model},''
	\href{http://dx.doi.org/10.1103/PhysRevD.86.034509}{{\em Phys. Rev. D}
		{\bfseries 86} (2012) 034509},
	\href{http://arxiv.org/abs/1203.0784}{{\ttfamily arXiv:1203.0784 [hep-lat]}}.
	
	\bibitem{Bazavov:2017dus}
	A.~Bazavov {\em et~al.}, ``{The QCD Equation of State to $\mathcal{O}(\mu_B^6)$
		from Lattice QCD},'' \href{http://dx.doi.org/10.1103/PhysRevD.95.054504}{{\em
			Phys. Rev. D} {\bfseries 95} no.~5, (2017) 054504},
	\href{http://arxiv.org/abs/1701.04325}{{\ttfamily arXiv:1701.04325
			[hep-lat]}}.
	
	\bibitem{Bollweg:2022rps}
	{\bfseries HotQCD} Collaboration, D.~Bollweg, J.~Goswami, O.~Kaczmarek,
	F.~Karsch, S.~Mukherjee, P.~Petreczky, C.~Schmidt, and P.~Scior, ``{Taylor
		expansions and Pad\'e approximants for cumulants of conserved charge
		fluctuations at nonvanishing chemical potentials},''
	\href{http://dx.doi.org/10.1103/PhysRevD.105.074511}{{\em Phys. Rev. D}
		{\bfseries 105} no.~7, (2022) 074511},
	\href{http://arxiv.org/abs/2202.09184}{{\ttfamily arXiv:2202.09184
			[hep-lat]}}.
	
	\bibitem{Borsanyi:2018grb}
	S.~Borsanyi, Z.~Fodor, J.~N. Guenther, S.~K. Katz, K.~K. Szabo, A.~Pasztor,
	I.~Portillo, and C.~Ratti, ``{Higher order fluctuations and correlations of
		conserved charges from lattice QCD},''
	\href{http://dx.doi.org/10.1007/JHEP10(2018)205}{{\em JHEP} {\bfseries 10}
		(2018) 205}, \href{http://arxiv.org/abs/1805.04445}{{\ttfamily
			arXiv:1805.04445 [hep-lat]}}.
	
	\bibitem{Gupta:2020pjd}
	S.~Gupta, D.~Mallick, D.~K. Mishra, B.~Mohanty, and N.~Xu, ``{Freeze-out and
		thermalization in relativistic heavy ion collisions},''
	\href{http://arxiv.org/abs/2004.04681}{{\ttfamily arXiv:2004.04681
			[hep-ph]}}.
	
	\bibitem{Xu:2016mqs}
	J.~Xu, ``{Energy Dependence of Moments of Net-Proton, Net-Kaon, and Net-Charge
		Multiplicity Distributions at STAR},''
	\href{http://dx.doi.org/10.1088/1742-6596/736/1/012002}{{\em J. Phys. Conf.
			Ser.} {\bfseries 736} no.~1, (2016) 012002},
	\href{http://arxiv.org/abs/1611.07134}{{\ttfamily arXiv:1611.07134
			[hep-ex]}}.
	
	\bibitem{Mukherjee:2023ijv}
	G.~Mukherjee, D.~Dutta, and D.~K. Mishra, ``{Conserved number fluctuations
		under global rotation in a hadron resonance gas model},''
	\href{http://arxiv.org/abs/2304.14658}{{\ttfamily arXiv:2304.14658
			[hep-ph]}}.
	
	\bibitem{Andronic:2009jd}
	A.~Andronic, P.~Braun-Munzinger, and J.~Stachel, ``{The Horn, the hadron mass
		spectrum and the QCD phase diagram: The Statistical model of hadron
		production in central nucleus-nucleus collisions},''
	\href{http://dx.doi.org/10.1016/j.nuclphysa.2009.12.048}{{\em Nucl. Phys. A}
		{\bfseries 834} (2010) 237C--240C},
	\href{http://arxiv.org/abs/0911.4931}{{\ttfamily arXiv:0911.4931 [nucl-th]}}.
	
	\bibitem{Andronic:2017pug}
	A.~Andronic, P.~Braun-Munzinger, K.~Redlich, and J.~Stachel, ``{Decoding the
		phase structure of QCD via particle production at high energy},''
	\href{http://dx.doi.org/10.1038/s41586-018-0491-6}{{\em Nature} {\bfseries
			561} no.~7723, (2018) 321--330},
	\href{http://arxiv.org/abs/1710.09425}{{\ttfamily arXiv:1710.09425
			[nucl-th]}}.
	
	\bibitem{Ko:2023whx}
	H.~S. Ko, ``{Higher-order moments of net-proton and the QCD phase structure},''
	\href{http://dx.doi.org/10.1051/epjconf/202327601019}{{\em EPJ Web Conf.}
		{\bfseries 276} (2023) 01019}.
	
	\bibitem{Rougemont:2015oea}
	R.~Rougemont, R.~Critelli, and J.~Noronha, ``{Holographic calculation of the
		QCD crossover temperature in a magnetic field},''
	\href{http://dx.doi.org/10.1103/PhysRevD.93.045013}{{\em Phys. Rev. D}
		{\bfseries 93} no.~4, (2016) 045013},
	\href{http://arxiv.org/abs/1505.07894}{{\ttfamily arXiv:1505.07894
			[hep-th]}}.
	
	\bibitem{Finazzo:2016mhm}
	S.~I. Finazzo, R.~Critelli, R.~Rougemont, and J.~Noronha, ``{Momentum transport
		in strongly coupled anisotropic plasmas in the presence of strong magnetic
		fields},'' \href{http://dx.doi.org/10.1103/PhysRevD.94.054020}{{\em Phys.
			Rev. D} {\bfseries 94} no.~5, (2016) 054020},
	\href{http://arxiv.org/abs/1605.06061}{{\ttfamily arXiv:1605.06061
			[hep-ph]}}. [Erratum: Phys.Rev.D 96, 019903 (2017)].
	
	\bibitem{Critelli:2016fvr}
	R.~Critelli, ``{Equilibrium and non-equilibrium properties of the QGP in a
		magnetic field: a holographic approach},''
	\href{http://dx.doi.org/10.1088/1742-6596/832/1/012047}{{\em J. Phys. Conf.
			Ser.} {\bfseries 832} no.~1, (2017) 012047},
	\href{http://arxiv.org/abs/1610.08751}{{\ttfamily arXiv:1610.08751
			[nucl-th]}}.
	
	\bibitem{Chen:2020ath}
	X.~Chen, L.~Zhang, D.~Li, D.~Hou, and M.~Huang, ``{Gluodynamics and
		deconfinement phase transition under rotation from holography},''
	\href{http://dx.doi.org/10.1007/JHEP07(2021)132}{{\em JHEP} {\bfseries 07}
		(2021) 132}, \href{http://arxiv.org/abs/2010.14478}{{\ttfamily
			arXiv:2010.14478 [hep-ph]}}.
	
	\bibitem{Fang:2021ucy}
	Z.~Fang, Y.-Y. Li, and Y.-L. Wu, ``{QCD phase diagram with a background
		magnetic field in an improved soft-wall AdS/QCD model},''
	\href{http://dx.doi.org/10.1140/epjc/s10052-021-09311-5}{{\em Eur. Phys. J.
			C} {\bfseries 81} no.~6, (2021) 545}.
	
	\bibitem{Zhao:2022uxc}
	Y.-Q. Zhao, S.~He, D.~Hou, L.~Li, and Z.~Li, ``{Phase diagram of holographic
		thermal dense QCD matter with rotation},''
	\href{http://dx.doi.org/10.1007/JHEP04(2023)115}{{\em JHEP} {\bfseries 04}
		(2023) 115}, \href{http://arxiv.org/abs/2212.14662}{{\ttfamily
			arXiv:2212.14662 [hep-ph]}}.
	
\end{thebibliography}
\providecommand{\href}[2]{#2}\begingroup\raggedright\endgroup

\end{document}